\documentclass[a4paper]{aa} 
\usepackage{amsmath} 
\usepackage{txfonts}
\usepackage{graphicx} 
\usepackage{url}
\usepackage{natbib} 
\usepackage{txfonts}
\usepackage[normalem]{ulem}
\usepackage{float}
\usepackage[caption = false]{subfig}
\bibpunct{(}{)}{;}{a}{}{,}

\newcommand{\Ms}{\ensuremath{M_\odot}}
\newcommand{\Zs}{\ensuremath{Z_\odot}}
\newcommand{\el}[2]{$\rm{}^{#2}\kern-0.6pt#1$}

\begin{document}

\title{The magnetic strip(s) in the advanced phases of stellar evolution}
\subtitle{Theoretical convective turnover timescale and Rossby number for  
low- and intermediate-mass stars up to the AGB at various metallicities}

\author{C. Charbonnel\inst{1,2} \and T. Decressin\inst{3} \and N.
Lagarde\inst{4} \and F. Gallet\inst{1}  \and A.Palacios\inst{5} \and
 M.Auri\`ere\inst{6} \and
R.Konstantinova-Antova\inst{7,2}  \and S.Mathis\inst{8}  \and R.I.~Anderson\inst{9,10} \and B.Dintrans\inst{2}}

\offprints{C.Charbonnel,\\ email: Corinne.Charbonnel@unige.ch}

\institute{$^1$ Department of Astronomy, University of Geneva, Chemin des
  Maillettes 51, 1290 Versoix, Switzerland \\
  $^2$ IRAP, UMR 5277, CNRS and Universit\'e de Toulouse, 14, av.E.Belin, 31400 Toulouse, France \\
  $^3$ INAF – Osservatorio Astronomico di Roma, via di Frascati 33, 00040 Monteporzio, Italy\\
  $^4$  Institut Utinam, CNRS UMR6213, Univ. Bourgogne Franche-Comt\'e, OSU THETA Franche-Comt\'e-Bourgogne, Observatoire de Besanon, BP 1615, 25010 Besanon Cedex, France \\
  $^5$ LUPM, UMR5299, Universit\'e de Montpellier, CNRS, place Eugne Bataillon, 34095 Montpellier, France \\
  $^6$  IRAP, UMR 5277, CNRS and Universit\'e de Toulouse, 57, av. d'Azereix, BP 826, 65008 Tarbes cedex, France \\
  $^7$  Institute of Astronomy and NAO, Bulgarian Academy of Sciences, 72 Tsarigradsko shose, 1784 Sofia, Bulgaria \\
   $^8$ Laboratoire AIM Paris-Saclay, CEA/Irfu Universit\'e Paris-Diderot CNRS/INSU, 91191 Gif-sur-Yvette, France \\
  $^9$  Physics and Astronomy Department, Johns Hopkins University, Baltimore, MD 21218, USA \\
  $^{10}$  Swis¥s National Science Foundation Fellow}
\date{A\&A accepted}

\authorrunning{} \titlerunning{The magnetic strip(s) in the advanced phases of stellar evolution}

\abstract{Recent spectropolarimetric observations of otherwise ordinary 
(in terms e.g. of surface rotation and chemical properties) G, K, and M giants have revealed localized magnetic strips in the Hertzsprung-Russell diagram coincident with the regions where the first dredge-up and core helium burning occur.}%
{We seek to understand the origin of magnetic fields in such late-type giant stars, which is currently unexplained. In analogy with late-type dwarf stars, we focus primarily on   parameters known to influence the generation of magnetic fields in the outer convective envelope. }
{We compute the classical dynamo parameters along the evolutionary tracks of low- and intermediate-mass stars at various metallicities using stellar models that have been extensively tested by spectroscopic and asteroseismic observations. Specifically, these include convective turnover timescales and convective Rossby numbers, computed from the pre-main sequence (PMS) to the tip of the red giant branch (RGB) or the early asymptotic giant branch (AGB) phase. 
To investigate the effects of the very extended outer convective envelope, we compute these parameters both for the entire convective envelope and locally, that is, at different depths within the envelope. We also compute the turnover timescales and corresponding Rossby numbers for the convective cores of intermediate-mass stars on the main sequence.} 
{Our models show that the Rossby number of the convective envelope becomes lower than unity in the well-delimited locations of the Hertzsprung-Russell diagram where magnetic fields have indeed been detected. }
{We show that $\alpha-\Omega$ dynamo processes might not be continuously operating, but that they are favored in the stellar convective envelope at two specific moments along the evolution tracks, that is, during the first dredge-up at the base of the RGB and during central helium burning in the helium-burning phase and early-AGB.
This general behavior can explain the so-called magnetic strips recently discovered by dedicated spectropolarimetric surveys of evolved stars.} %

\keywords{Dynamo - Stars: Activity - Interiors - Magnetic fields - Rotation}

\maketitle

\section{Introduction}

Magnetic fields are actively searched for at the surface of all kinds of stars throughout the Hertzsprung-Russell  diagram (HRD), as they probably impact stellar evolution from birth to death in various ways (see e.g., \citealt{DonatiLandstreet09} for a general review, and \citealt{Wadeetal16}, \citealt{Alecianetal13}, \citealt{Vidottoetal14}, \citealt{Folsometal16} for recent spectropolarimetric surveys on massive, intermediate-mass, and low-mass stars).
Recently, magnetic fields have been unambiguously detected via Zeeman signatures in a large sample of single G-K giants observed with the spectropolatimeters Narval@TBL and ESPaDOnS@CFHT \citep[][]{Konstantinovaetal2013,Auriereetal2015,Borisovaetal16,Tsvetkovaetal16arXiv}.
Interestingly, the cool intermediate-mass evolved stars with surface magnetic fields are found to cluster in certain regions of the HRD that correspond to precise moments of their evolution where convective envelopes make up a significant fraction of the stellar mass. 
These observations are of particular importance for stellar-evolution modeling, since magnetic fields may play a crucial role in the angular momentum evolution of different types of stars. 
For instance, it is well established that magnetic fields affect the rotation rate of solar-type
stars through the torque applied to stellar surfaces by magnetically coupled stellar winds \citep[e.g.,][]{Schatzman1962,Mattetal2015,Revilleetal15,Amardetal16}. 
However, the efficiency and the impact of magnetic braking for evolved stars  
are not well studied yet. In addition, because of the angular momentum they transport \citep[e.g.,][]{Spruit1999,MaZa05}, magnetic fields may play an important role in explaining the properties of core and surface rotation in giants as seen by asteroseismology \citep[e.g.,][]{Beck2012,Mosser2012a,Mosser2012b,Deheuvels2012,Deheuvelsetal2014,Cantiello2014,DiMauro2016}, as well as the rotation rate of white dwarf remnants \citep[][]{Suijsetal2008}.
By comparing the observed rotational and magnetic properties of stars at
different evolutionary phases with the predictions of rotating stellar
models, one may thus obtain strong constraints on the input physics of the stellar models. 
Conversely, it is important to know whether the possible presence and the global properties of the magnetic field of a given star can be anticipated from its position in the Hertzsprung-Russell diagram (see \citealt{Gregoryetal12} for the case of pre-main sequence stars). 

In main sequence solar-like stars and their red giant descendants, dynamo processes are invoked to generate
magnetic fields through turbulence and rotational shear within the stellar
convective envelopes or within a thin shear layer at the interface between the
convective and the radiative regions 
 \citep[so-called tachocline; see e.g.,][ for a review and references therein]{CharbonneauARAA2014}. 
 This is supported by the existence of a tight correlation between the measured magnetic field strength and the rotational properties (namely the rotational period and the convective Rossby number Ro, hereafter Rossby number for simplicity) of magnetic red giant stars with known rotational periods \citep{Auriereetal2015}.
In hotter, more massive main sequence and Ap stars that do not have extended convective envelopes, fossil magnetic fields are generally invoked to explain magnetic field properties  \citep[e.g.,][]{Neineretal2015}.
Moreover, those early-type stars might produce internal magnetic fields in their vigorous convective core during the central hydrogen-burning phase \citep[e.g.,][]{Brunetal2005,Stelloetal2016Nature}.
When these objects move towards the red giant branch (RGB), deeply buried magnetic fields might survive and interact with the deepening convective envelope.
Some post-main sequence stars have been recently shown to be possible Ap star descendants \citep[see e.g., ][ and references
therein]{Auriereetal_proceedings2014}. 
However, the details on how dynamos and field dredge-up can operate in giant stars are not yet sufficiently understood. 

The properties of the convective regions as well as the rotation velocity of
stars depend on their initial mass and metallicity, and strongly vary along
their evolution.
As a consequence, the efficiency of the dynamo-driven mechanisms is expected to vary  
as stars of various initial masses evolve. 
Therefore, it is necessary to investigate how the main stellar quantities that are currently used to evaluate the efficiency
of magnetic field generation in stellar interiors do change as stars of different initial masses evolve.
In this context, the classical relevant quantities are 
the convective turnover timescale, which varies with spectral type as a function of the evolving properties of the stellar convective zones, 
and the Rossby number, which quantifies the interactions between
convection and rotation.
This latest quantity is proportional to the ratio between the stellar rotation
period and the convective turnover time,  and it describes how the Coriolis
force affects convective eddies. Low Rossby values (of the order or below unity) are 
expected in the case of stars with sufficiently fast rotation for efficient magnetic field generation
\citep[e.g.,][]{DurneyLatour1978,MangeneyPraderie1984,Noyesetal1984,Brunetal2015,Augustsonetal16}.

The present paper describes how these classical dynamo parameters vary throughout the evolution of
low- and intermediate-mass stars at various metallicities, based on the predictions on one-dimensional (1D) stellar models. 
This allows us to identify the evolution phases during which stellar dynamo may operate 
and to motivate further studies involving detailed, sophisticated, and numerically expensive, magnetohydrodynamic simulations that are needed to capture all the processes involved in global stellar dynamos (e.g., \citealt{Brun2015}, and references therein). Here we use the grid of stellar models
presented in \citet{Lagardeetal12}, whose
characteristics are briefly recalled in \S~\ref{Section:DetailsGrid}; we also describe the content of the electronic tables. 
In \S~\ref{Section:Tauconv}, we present convective turnover timescales
from the PMS to the early Asymptotic Giant Phase (AGB) and along
the thermally pulsing AGB (TP-AGB) for selected cases, for the convective envelope of stars
with initial masses between 0.85 and 6.0 M$_{\odot}$, and for two metallicities Z = 0.014 and Z = 0.0001.
We discuss 
how the turnover timescale varies with depth inside the
stellar convective envelope as stars evolve, and explore how this depends on
stellar mass and metallicity. 
In \S~\ref{Section:Rossby} we give the theoretical Rossby
numbers  for our stellar models including rotation and discuss their variations from the zero-age main sequence (ZAMS) to the early-AGB. 
We present theoretical evidence for the existence of two magnetic strips that low- and intermediate-mass stars are predicted to cross  
in the advanced phases of their evolution. In \S~\ref{section:comparison} we compare our model predictions with the position in the HRD of  red giant stars for which surface magnetic fields have been detected. We summarize our results in \S~\ref{conclusion}. In the Appendix, we provide the turnover timescales and the Rossby numbers for the convective core of intermediate-mass main sequence stars for the solar metallicity grid.  

\section{Reference grid of stellar models and content of the electronic tables}
\label{Section:DetailsGrid}

This paper is based on the grid of standard and rotating stellar models computed
 by \citet{Lagardeetal12}\footnote{http://cdsarc.u-strasbg.fr/viz-bin/qcat?J/A+A/543/A108} with the code STAREVOL (V3.00) for a range of initial
masses between 0.85 and 6~M$_{\odot}$ and for four values of [Fe/H] = 0, -0.56,
-0.86, -2.16 (corresponding respectively to metallicities
Z=0.014\footnote{Z=0.014 corresponds to Z$_{\odot}$, with \citet{Asplund05}
chemical composition except for Ne for which we use the value derived by
\citet{Cunha06}.}, 4$\times 10^{-3}$, 2 $\times 10^{-3}$, and $10^{-4}$).
Convection is treated according to the classical mixing length formalism with $\alpha_{MLT}$=1.6. The boundary between convective and radiative layers is defined with the Schwarzschild criterion. An overshoot parameter d/H$_p$ is taken into account for the convective
core; it is set to 0.05 for stars with less than 2.0~M$_{\odot}$ and to 0.10  for
more massive stars. 
The evolution of angular momentum within the radiative stellar regions is
computed using the complete formalism developed by \citet {Zahn92}, \citet
{MaeZah98}, and \citet{MaZa04}, which takes into account advection by meridional
circulation and diffusion by shear turbulence (see
\citealt{Palacios03,Palacios06,Decressin09}). Convective regions are treated as solid bodies.
The impact of rotation-induced mixing and thermohaline instability on the global parameters and the 
chemical properties of low- and intermediate-mass stars is discussed in
\citet{ChaLag10}.
The corresponding nucleosynthetic yields are presented in \citet{Lagardeetal11II}.
The relevant classical stellar parameters as well as the theoretical global
asteroseismic properties of the models can be found in \citet{Lagardeetal12}.

The initial rotation velocity of the models corresponds to $30 \%$ of the
critical velocity at the ZAMS \citep{Lagardeetal2014corrigendum}.
For the 2~M$_{\odot}$, Z$_{\odot}$ case, three additional rotating models are
presented here, with initial rotation velocity on the ZAMS of 50, 163, and 250
km.sec$^{-1}$, which correspond to 15, 44, and 68$ \%$ of the critical
velocity.
Magnetic braking is included following \citet{Kawaleretal88} from
the ZAMS onward for all low-mass models, that is, with 1.25~M$_{\odot}$,
Z$_{\odot}$, and 0.85~M$_{\odot}$, Z=0.0001. 
For more massive stars that have very thin convective envelope on the main sequence, the evolution of the surface velocity is  
governed by secular effects and internal transport of
angular momentum through meridional circulation and shear turbulence.
Rotation is accounted for throughout the evolution from
the ZAMS up to the beginning of the TP-AGB phase for all the models that 
ignite central helium burning in non-degenerate conditions; for the less massive
models, rotation is included only until the tip of the red giant branch.

\begin{figure}[!b]
\centering
 \includegraphics[width=0.4\textwidth, angle=0]{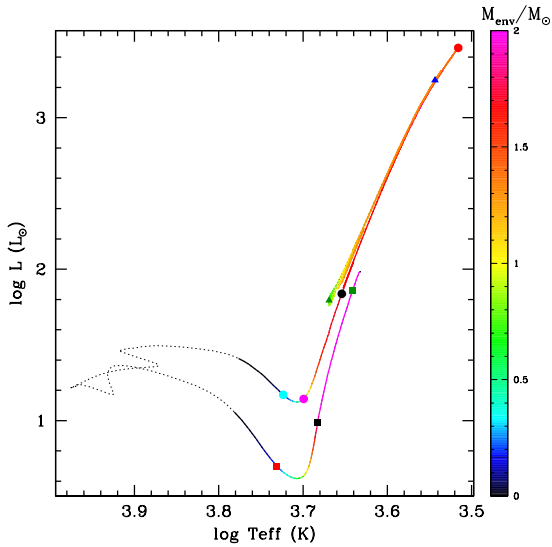}
 \caption{Variations of the size of the convective envelope (in solar mass, color-coded) along the track of the 2~M$_{\odot}$, Z$_{\odot}$ standard model. 
 The dotted line corresponds to the phases when the convective envelope contains less than
 0.001~M$_{\odot}$. Specific evolution points are selected. On the PMS, the green, black, and red squares correspond respectively to stages
 when the star is still fully convective, and when the mass of the receding
 convective envelope is 1.9 and 0.26~M$_{\odot}$. In the advanced phases, the
 magenta, cyan, black circles correspond respectively to the beginning,
 middle, and end of the first dredge-up, that is, when the mass of the convective
 envelope is 0.13, 0.87, and 1.7~M$_{\odot}$. Red circle, green and blue triangles
 correspond to the RGB tip, to the middle of central helium-burning, and to the
 end of the second dredge-up on the early-AGB, respectively; at these evolutionary points the
 mass of the convective envelope is respectively 1.39, 0.7, and 1.31~M$_{\odot}$}            
 \label{fig:HRD2Msun_Menv}
\end{figure}

We provide electronic files containing sets of relevant physical quantities tabulated for 500 evolution points along the evolution tracks from the
PMS up to the early-AGB phase \citep[see \S~3.1 in][]{Lagardeetal12}. 
For each model, we give the turnover timescales at different heights in the convective envelope $ \tau_{H_p /2}$, $\tau_{H_p}$, $\tau_{R_{EC}/2}$,
$\tau_{M_{EC}/2}$, and $\tau_{max}$, as well as the corresponding Rossby numbers (see \S~3 and 4) in addition to the quantities already listed in Table~2 of \citet{Lagardeetal12}.

\section{Convective turnover timescales} 
\label{Section:Tauconv}

\subsection{Definitions}

\begin{figure*}
\centering
 \includegraphics[width=0.8\textwidth, angle=0]{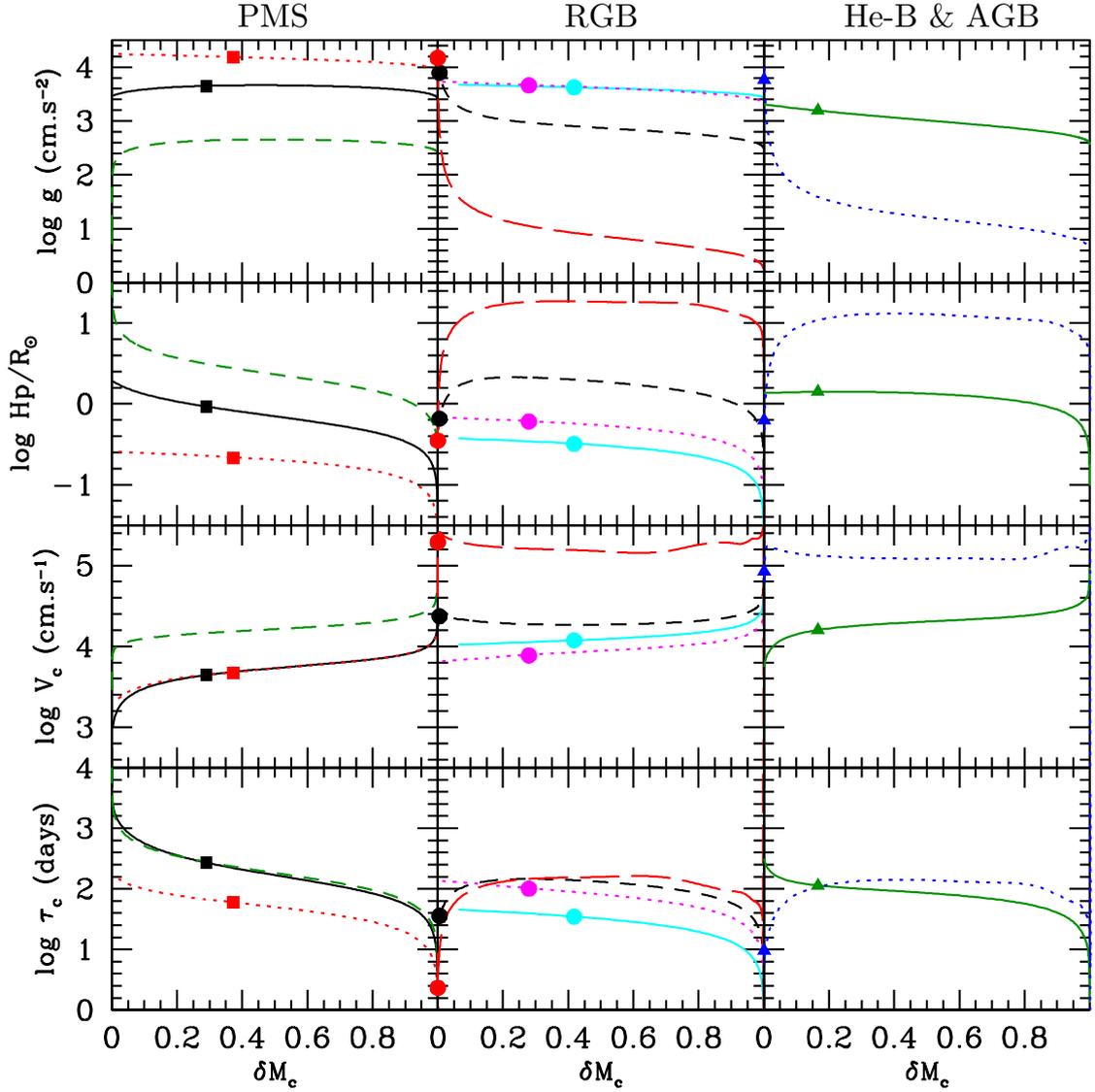} 
   \caption{Variations within the convective envelope of gravity, pressure scale height, convective velocity, and turnover time (from top to bottom) for the 2~M$_{\odot}$, Z$_{\odot}$ standard model. The abscissa is the scaled mass coordinate $\delta M_{conv}$ ($\delta M_{conv}$=0 and 1 respectively at the base and at the top of the convective envelope). The colors correspond to evolution points selected in Fig.~\ref{fig:HRD2Msun_Menv} (see there for the corresponding values of the envelope thickness). PMS, RGB, and advanced phases are shown on the left, middle, and right panels, respectively. The corresponding colored symbols indicate the location of H$_p$/2 above the base of the convective envelope (for the full convective model on the PMS - green line - this point would be located outside the star, at a radius of $\sim 26$ R$_*$)
   }
  \label{fig:tconvprofiles2Msun}
\end{figure*}

The local convective turnover time at a given radius $r$ inside the convective
envelope, $\tau_{c}(r)$ and the  so-called global convective turnover time,
$\tau_g$ are defined as in 
\citet{Gilliland85}
\begin{equation}
\tau_{c}(r)= \alpha_{MLT} H_P (r) /V_c(r),
\end{equation}
and  
\begin{equation}
\tau_g = \int_{R_b}^{R_t} \frac{dr}{V_c (r)}.
\end{equation}
$H_P(r)$ is the local pressure scale height,  
$\alpha_{MLT}$ is the mixing-length parameter, $R_b$ and $R_t$ are the radii at the bottom and at the top of the
convective envelope, respectively (taken where $\bigtriangledown - \bigtriangledown_{ad} =0$,
following Schwarzschild criterion). 
According to the mixing length theory the average local convective velocity is defined as 
\begin{equation} 
\label{eqVC}
V_c^2(r) = g \delta (\nabla -
\nabla_{ad}) \frac{\alpha_{MLT}^2}{8 H_P(r)}, 
\end{equation}
where $g$ is gravity and $\delta = \partial \ln \rho/\partial \ln T$
is determined from the equation of state.

Since $\tau_{c}(r)$ is strongly depth-dependent (see \S~\ref{subsection:2M}), and since it is not well established where exactly a dynamo operates within the
stellar convective envelope, we determine its value at different depths within the stellar convective envelope. 
$\tau_{H_p /2}$ and $\tau_{H_p}$ are the convective turnover times at
half a pressure scale height and at a pressure scale height above the base of
the convective envelope, respectively.
$\tau_{R_{EC}/2}$ and $\tau_{M_{EC}/2}$ are computed at half of the radius and
half of the mass of the convective envelope, and $\tau_{max}$ is the maximum
convective turnover time within the convective envelope. We investigate below how these quantities vary from the PMS along the Hayashi track until the end of the early-AGB for all models included in the grid that comprises both standard models and models including rotation, for four different metallicities.
We provide a detailed discussion of the  2~$M_{\odot}$, Z$_{\odot}$ standard model in \S~\ref{subsection:2M} and then describe the trends
with mass and metallicity in \S~\ref{subsection:mass}.
We also compute the turnover timescales in the convective core of solar metallicity main sequence stars (when this core is present), using the same definitions as above, and present the results in the Appendix. This allows us to provide complete diagnoses for the studied stars.

 \begin{figure*}[!ht]
 \centering
 \includegraphics[width=0.35\textwidth, angle=-90]{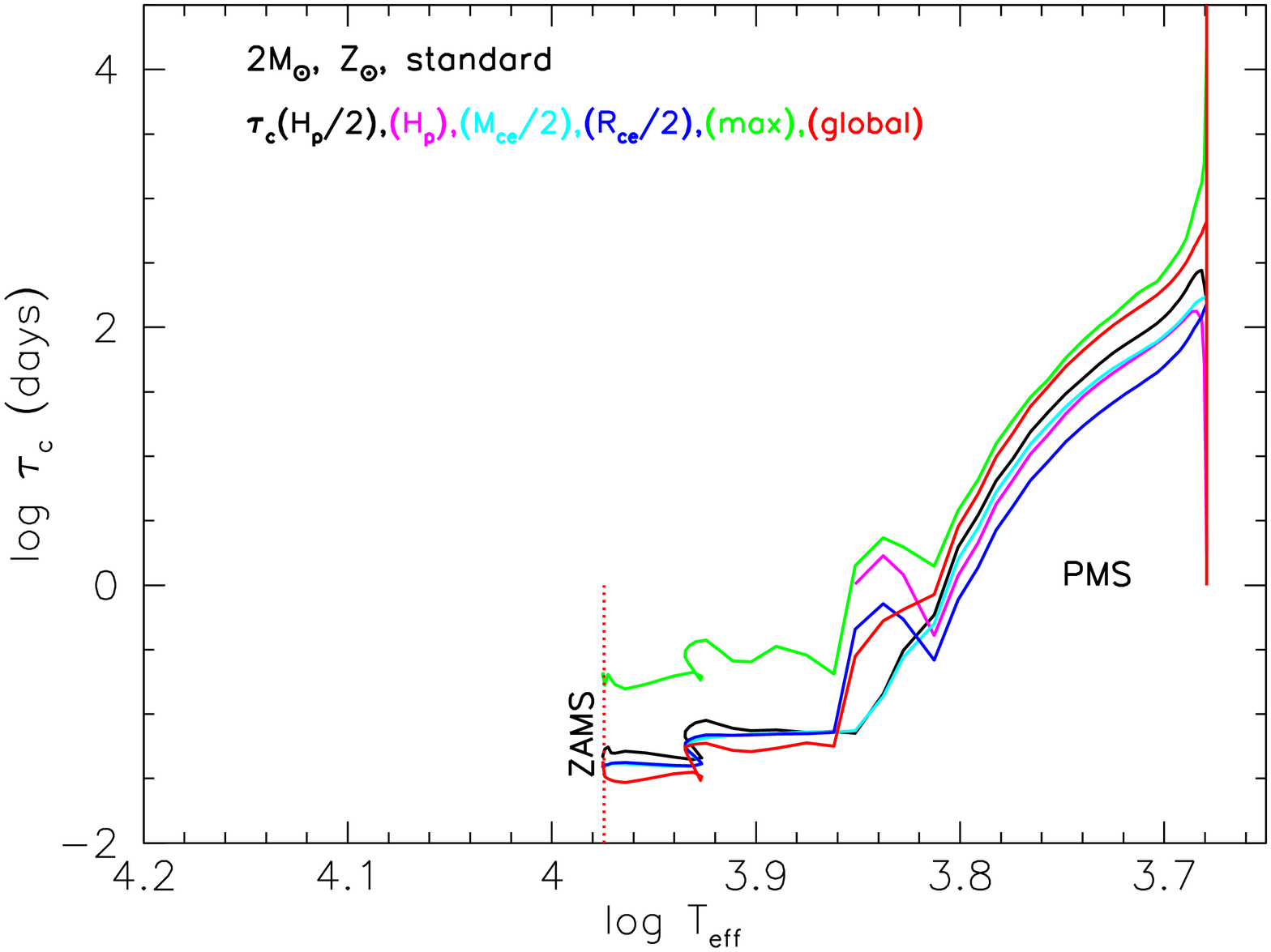} \includegraphics[width=0.35\textwidth, angle=-90]{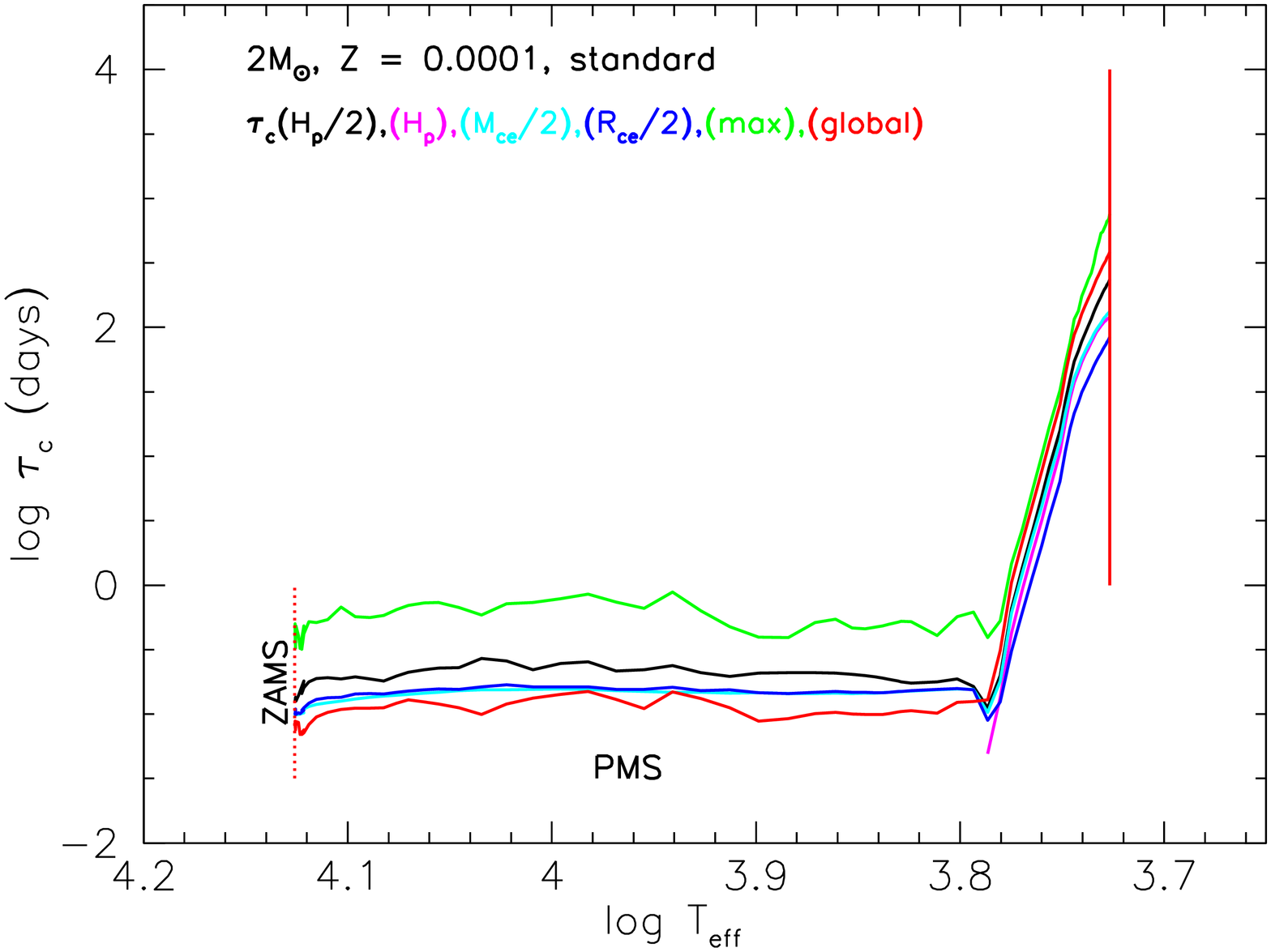}\\
  \includegraphics[width=0.35\textwidth, angle=-90]{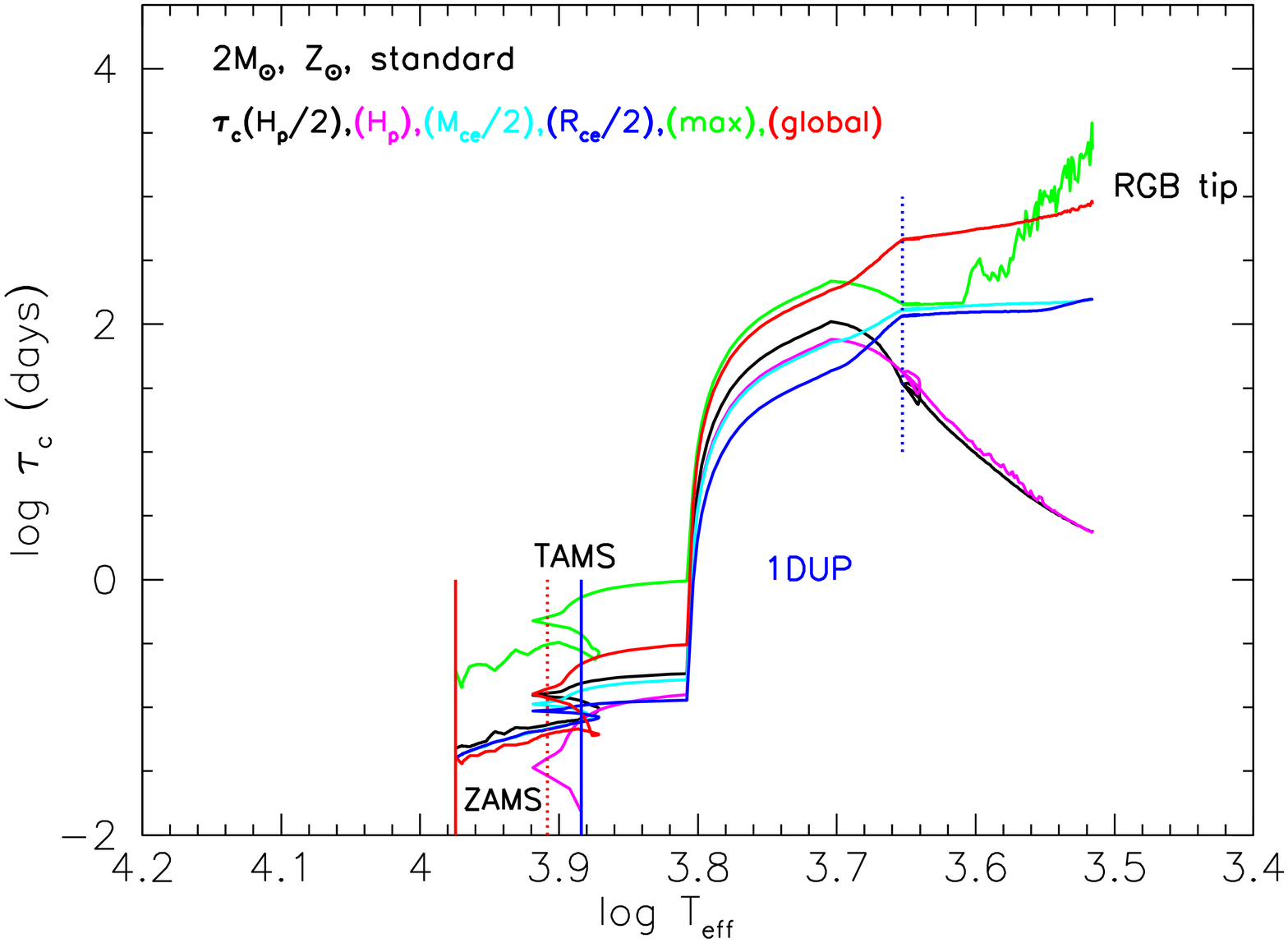} \includegraphics[width=0.35\textwidth, angle=-90]{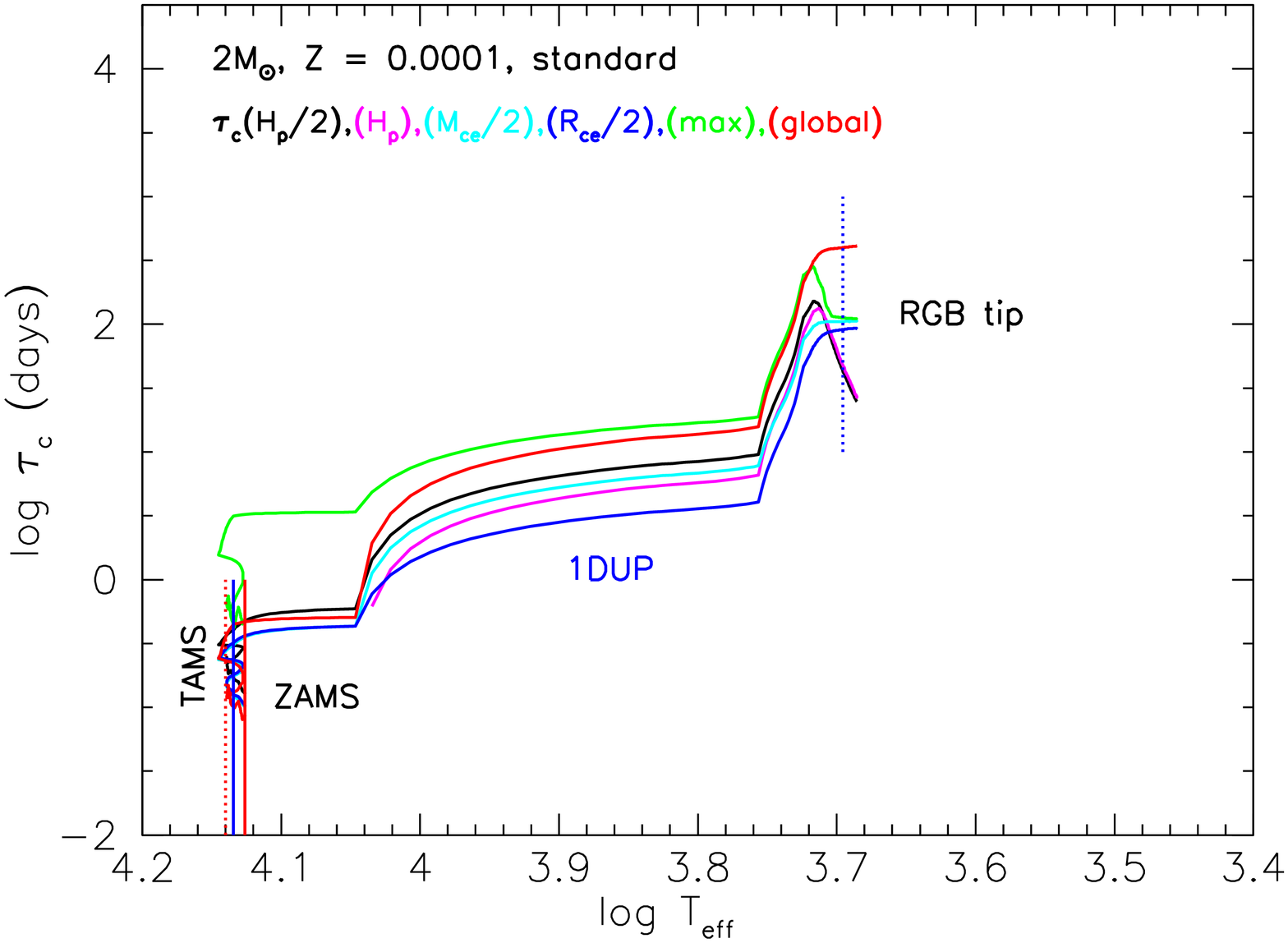}\\
  \includegraphics[width=0.35\textwidth, angle=-90]{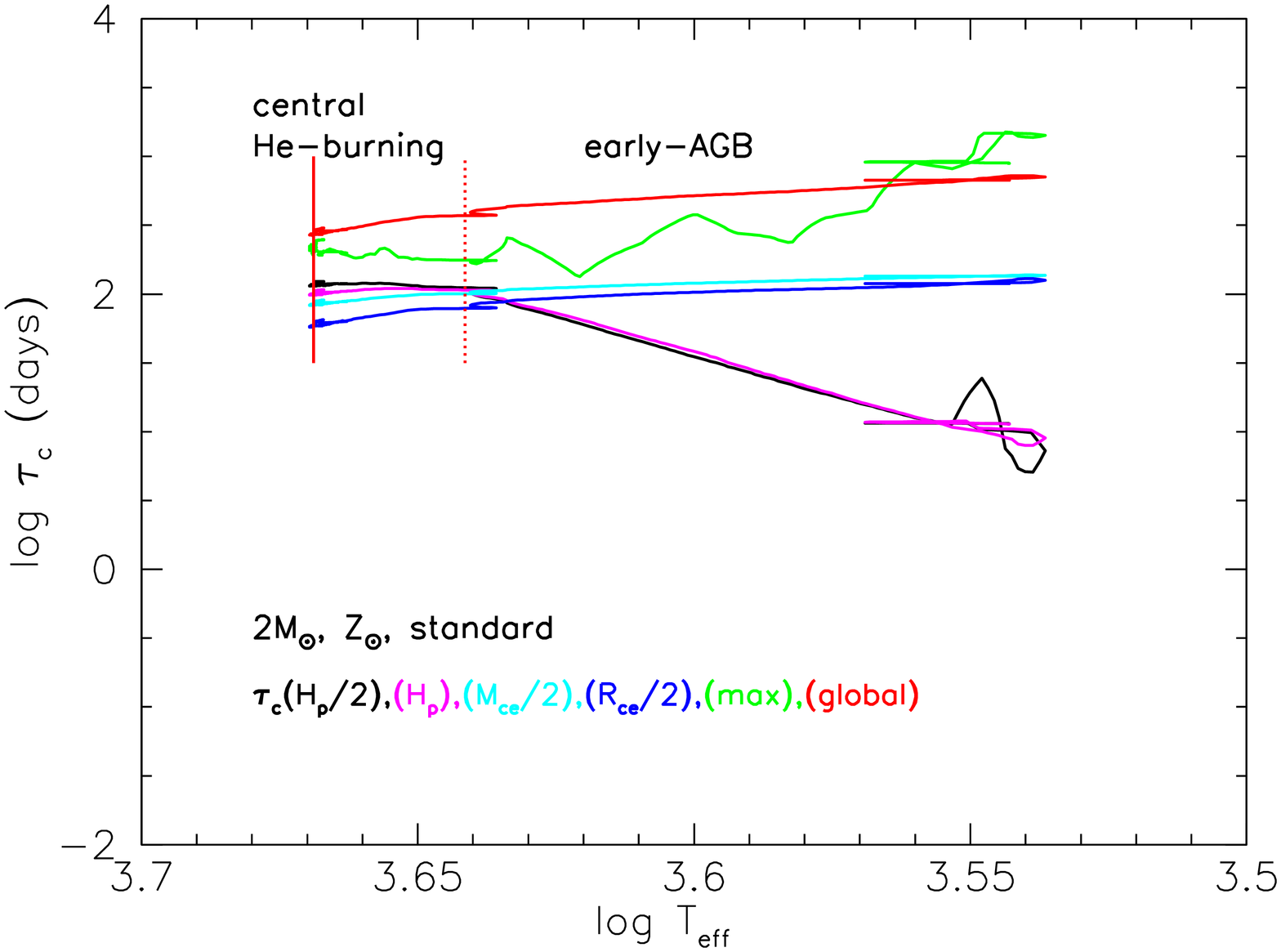} \includegraphics[width=0.35\textwidth, angle=-90]{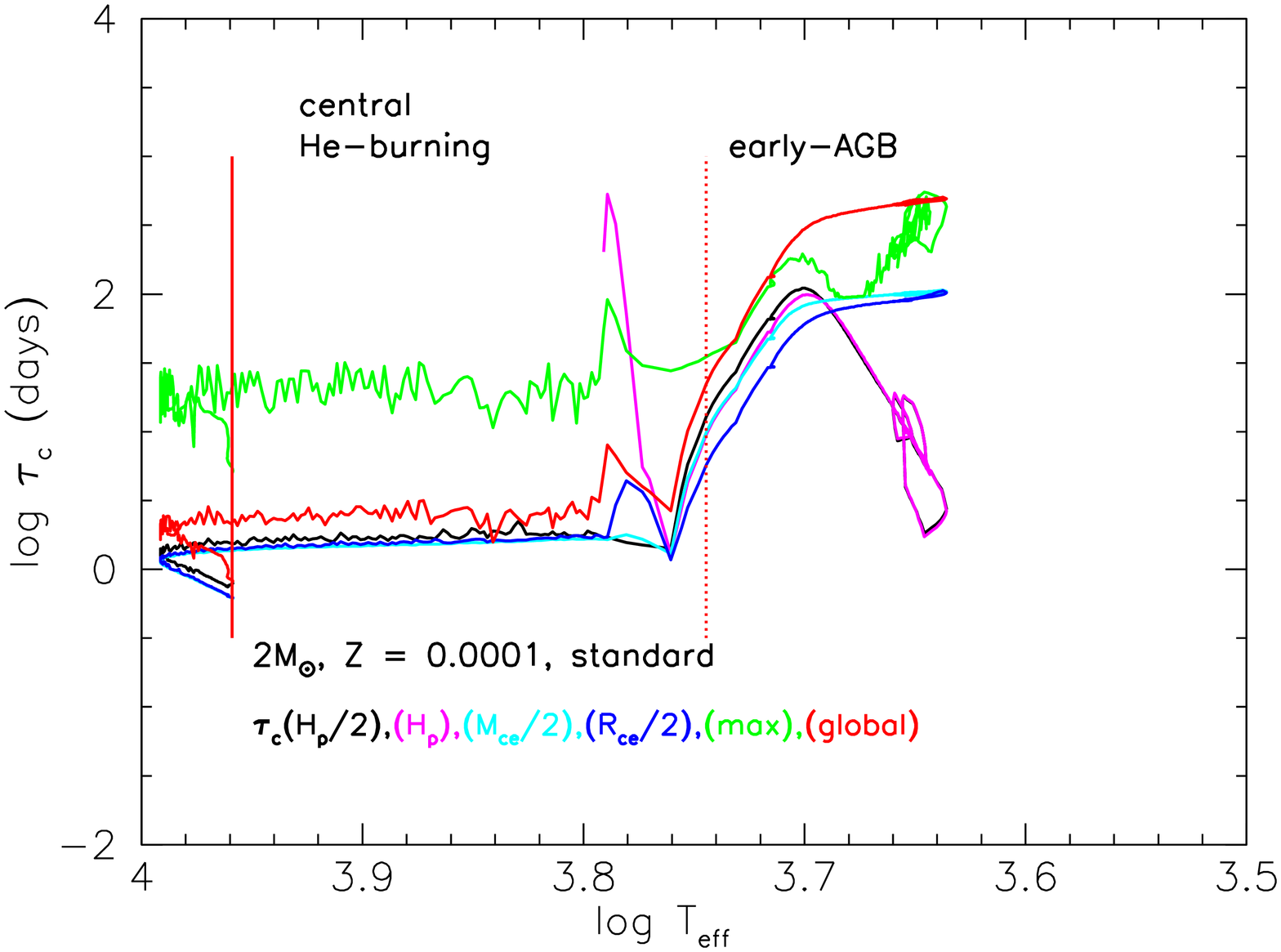}\\
   \caption{Evolution of the convective turnover time at different depths within the convective envelope as a function of effective temperature
     on the PMS (top), from the ZAMS up to the RGB tip (middle), and from the beginning of central helium-burning up to the end of the early-AGB (bottom). 
     The vertical lines indicate the beginning and the end (solid and dashed, respectively) of the PMS, of the central H- and He-burning phases, and of the first dredge-up (in blue).
     Left and right figures correspond to
     the 2~M$_{\odot}$ standard models computed at Z~$_{\odot}$ and
     Z=0.0001 respectively.   The magenta lines that are associated to $\tau_c(H_p)$ are sometimes interrupted; this corresponds to evolution phases when the radius at a pressure scale height $H_p$ above the base of the convective envelope becomes larger than the radius of the star}
  \label{fig:m2p0zsunZ0001}
\end{figure*}

\subsection{Turnover timescale in stellar convective envelopes along the evolution}
\label{subsection:CE}
\subsubsection{The case of the 2~M$_{\odot}$, Z$_{\odot}$ model}
\label{subsection:2M}

Fig.~\ref{fig:HRD2Msun_Menv}  shows the variations of the mass enclosed in the convective envelope along the evolution track in the Hertzsprung-Russell diagram in the case of the 2~$M_{\odot}$, Z$_{\odot}$ standard model.   
We select evolution points along the track for which we show in Fig.~\ref{fig:tconvprofiles2Msun} the profiles of the various quantities that enter in the
definition of the turnover timescale.
Finally, Fig.~\ref{fig:m2p0zsunZ0001} shows the variation of the convective turnover timescales at different depths
within the convective envelope ($\tau_{H_p /2}$, $\tau_{H_p}$, $\tau_{R_{EC}/2}$, $\tau_{M_{EC}/2}$, $\tau_{max}$, and $\tau_g$, in days,
logarithmic scale) throughout the evolution of the star.

{\bf Pre-main sequence --}
PMS evolution proceeds along increasing T$_{\rm eff}$. 
At first, the star is fully convective. As it contracts along the Hayashi track, the convective envelope quickly
decreases in mass. Gravity and $H_p$ strongly vary (increase
and decrease, respectively), which impacts $V_c$
following Eq.~\ref{eqVC} (Fig.~\ref{fig:tconvprofiles2Msun}).
Consequently the convective turnover timescale decreases during the contraction phase.
Then the base of the convective envelope as well as the various $\tau_{c}(r)$
stay relatively constant as the star evolves towards the ZAMS. This behavior has already been described in the literature (cf e.g., \citealt{KimDemarque1996} and \citealt{Landinetal10}). 
In our 2~$M_{\odot}$, Z$_{\odot}$, 
$\tau_{H_p /2}$ (respectively $\tau_{max}$) drops from 290 to 0.04 days
(respectively from 3150 years to 0.17 days) along the PMS.

{\bf Main sequence --} While on the MS, the external convective region of this model is extremely thin (less than 0.001~M$_{\odot}$), and it varies very little in mass. 
Therefore the various $\tau_{c}(r)$ have very low values, and they stay relatively constant as the star evolves towards cooler T$_{\rm eff}$ from the ZAMS to the Terminal Age Main Sequence (TAMS).

{\bf Hertzsprung gap and RGB --}
As the star adjusts to hydrogen-shell burning and moves towards cooler
T$_{\rm eff}$ along the subgiant branch (also called Hertzsprung gap), the
convective envelope strongly deepens, thereby increasing in mass (the
so-called first dredge-up, hereafter 1DUP), before receding again as the star
expands in radius along the RGB (see Fig.~\ref{fig:HRD2Msun_Menv}). Simultaneously,
within the bulk of the convective envelope the pressure scale height
strongly increases while gravity decreases by approximately three orders of magnitude
up to the RGB tip (Fig.~\ref{fig:tconvprofiles2Msun}). As a consequence, the
turnover timescale strongly rises at all depths within the convective
envelope during the 1DUP. In the upper part of the RGB (Fig. \ref{fig:m2p0zsunZ0001}),
$\tau_{H_p/2}$ and $\tau_{H_p}$ slightly decrease again, due to the pressure profile, whose
gradient then strongly steepens close to the base of the convective
envelope. The other $\tau_{c}$ all continue to increase up to the RGB tip
mainly due to the decrease in gravity. The increase in the value of
  $\tau_\text{max}$ in the upper part of the RGB is driven by the changes
  in opacity in the stellar atmosphere that weaken the convection velocity
near the surface.

{\bf Clump and early-AGB --}
When the star ignites central helium burning in the so-called clump, its
convective envelope is much thinner than at the RGB tip, and gravity has
increased again; the quantities associated to the different definitions of $\tau_{c}$ then stay relatively
constant.
As the star evolves towards the early-AGB, its convective envelope deepens
again. During this phase, $\tau_{c}$ increase again within the bulk of the convective envelope, except in the deepest convective
layers where the gradient of $H_P$ strongly steepens.

\subsubsection{Metallicity effect}

The question of whether one can use the same set of convective turnover times for stars of different populations was addressed by \cite{RucinskiVdB90}. However, their study was limited to the case of zero-age main-sequence solar-type stars with [Fe/H] above solar. Using our model grid, we can estimate the impact of large variations in metallicity on the convective turnover timescale over a wide range of evolutionary phases (from the PMS to the AGB).

A low-mass stellar model at Z=0.0001 evolves at higher luminosity than its Z$_{\odot}$ counterpart,  and it reaches
higher T$_{\rm eff}$ on the ZAMS. This results from the well-known impact of metals on the radiative opacity and on the mean molecular weight in stellar interiors \citep[e.g.,][]{Kippenhahnbook}.
This slightly modifies the convective
properties of the star, leading to higher values of $\tau_{c}(r)$ at a given
effective temperature during the evolution towards the ZAMS, as shown in
Fig.~\ref{fig:m2p0zsunZ0001} for the 2~M$_{\odot}$ case. For the same reason
(i.e. opacity effects), the different $\tau_{c}(r)$ are also higher for the
most metal-poor model on the MS, the subgiant phase, the RGB,  and the AGB.
During central helium-burning, the low-metallicity model evolves along
so-called blue loops, reaching much higher T$_{\rm eff}$
than the Z$_{\odot}$ model, which stays in the red clump. This results in
thinner convective envelope and lower $\tau_{c}$ at all depths.
Consequently, we can conclude that  sets of convective turnover times computed with the relevant metallicity should be used for stars of different populations.

\subsubsection{Mass dependence}\label{subsection:mass}

\begin{figure}
\centering
 \includegraphics[width=0.37\textwidth, angle=-90]{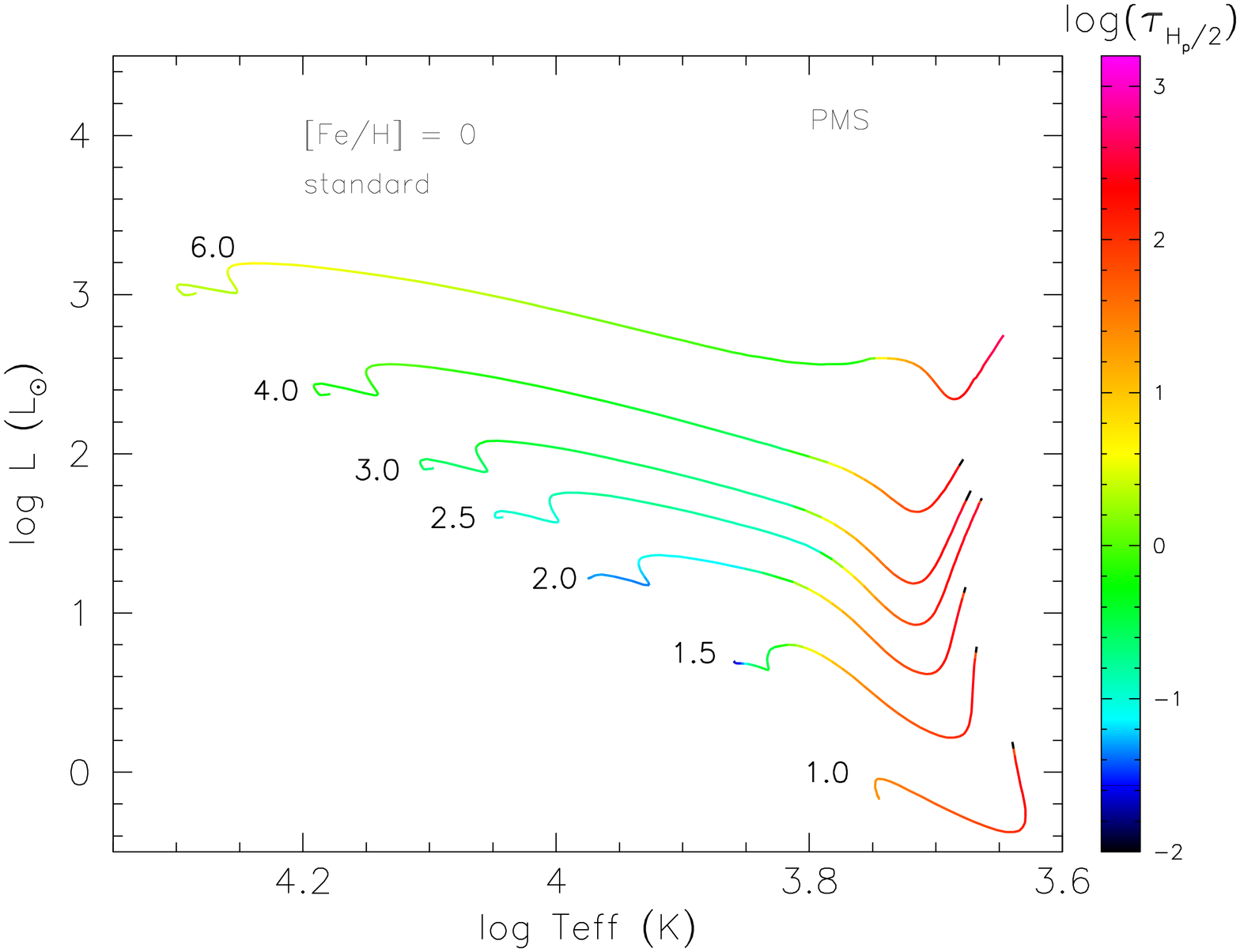}
  \includegraphics[width=0.37\textwidth, angle=-90]{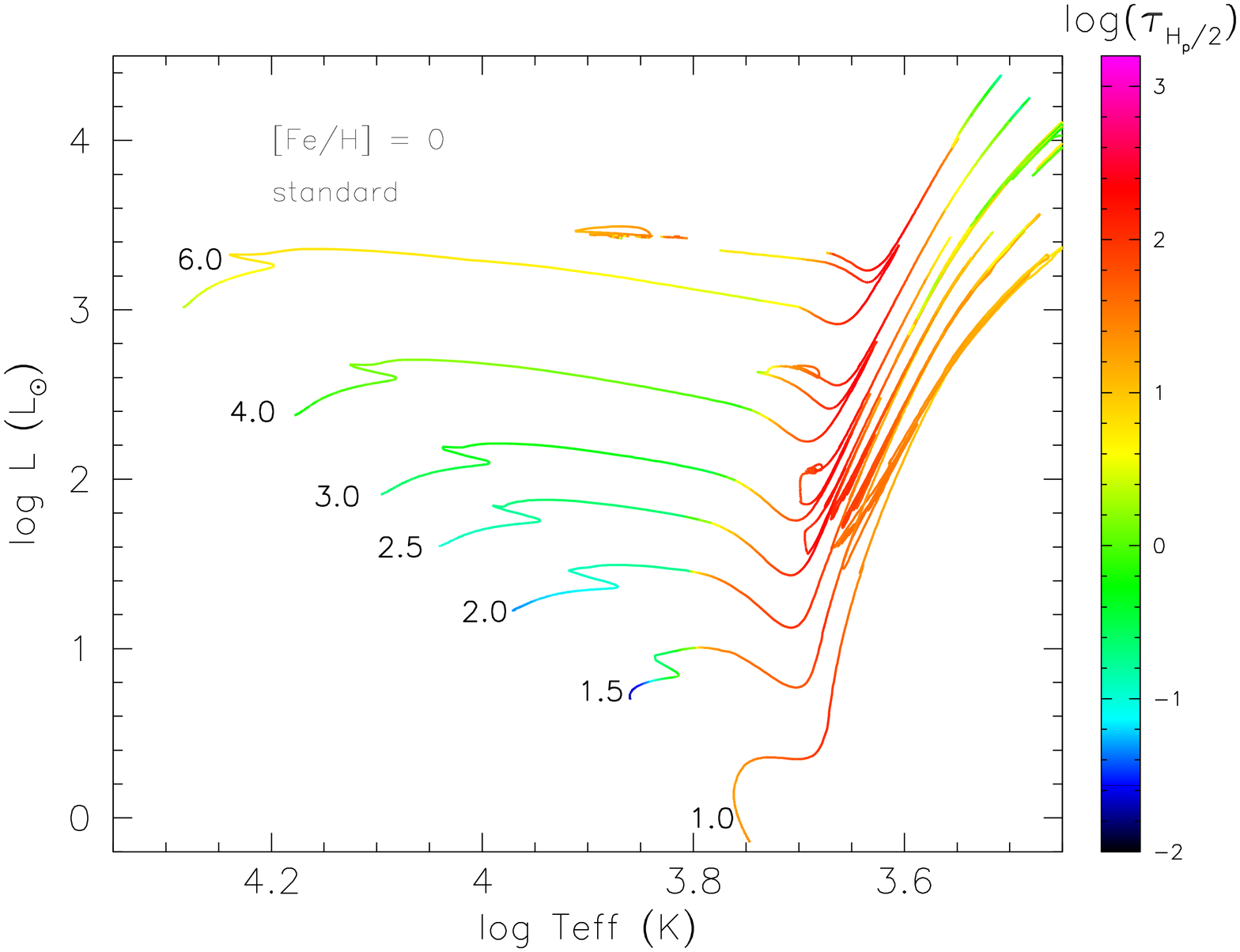} \hspace{0.cm}  
\caption{Color-coded convective turnover time $\tau_{H_p /2}$ (in logarithmic scale) in the stellar convective envelope along the PMS  and beyond the ZAMS (top and bottom panels respectively) for the 
solar metallicity models. Initial stellar masses (in solar mass) are indicated on the tracks}
  \label{fig:hrdtconvZsunstand}
\end{figure}

The general behavior of the convective turnover timescales at different depths within
the stellar convective envelope along the evolution tracks is similar for all stars to that described above for the 2~M$_{\odot}$ case. 
To illustrate the evolutionary aspects we chose to focus on $\tau_{H_p /2}$ which is shown in Fig.~\ref{fig:hrdtconvZsunstand} for all the solar metallicity models of our grid along the PMS and from the ZAMS onwards (top and bottom panels respectively). Importantly, we see 
that for all the models, $\tau_{H_p /2}$ sharply increases when the post-main sequence stars reach the middle of the Hertzsprung gap.  
In particular, $\tau_{H_p /2}$ becomes higher than approximately one day when log T$_{\rm{eff}}$ becomes lower than $\sim 3.8 - 3.7$ (see also Fig.~ \ref{fig:m2p0zsunZ0001}), which corresponds to the evolutionary point when the deepening convective envelope reaches a mass of $\sim$ 1 to 5$\%$ of the stellar mass. It is maximum (with very similar values over the whole mass range considered here) at the base of the RGB, when the convective envelope encompasses $\sim 40 \%$ of the total stellar mass.
Interestingly, the domain in effective temperature when the convective turnover times are the highest is very similar on the pre- and post-main sequences, as  
the fractional envelope mass M$_{env}$/M$_*$ of the pre- and post-main sequence stars are very similar at a given effective temperature. This proxy for internal structure is a direct reflection of the growth of the convective envelope in the stellar evolution phases when rapid stellar contraction or extension occur (pre- and post-main sequence respectively).  \citet{Gregoryetal12} suggested that it can be used as a proxy for the general characteristics of a magnetic field in PMS stars at a given position in the HRD. Here, we anticipate that it is also the case for post-main sequence stars.

\subsubsection{Impact of rotation}
We refer to \citet{ChaLag10} and \citet{Lagardeetal12} for a detailed discussion of the impact of rotation on the global parameters of our grid of models, and focus  only on the impact on the evolution paths in the HRD that are relevant for the convective properties of the stars. 
For a given stellar mass and metallicity, the evolutionary effects of rotation shift the evolution tracks to slightly higher luminosity. However, for a given T$_{\rm{eff}}$ 
the properties of the convective envelope change very little, and one finds very similar $\tau_{H_p /2}$
values (the same is true for all $\tau_{c}(r)$).   
Therefore, considering our current modeling of the impact of rotation on stellar evolution one can conclude that for a given stellar mass the turnover
timescales do not significantly depend on initial rotation velocity. 

\section{Rossby number}
\label{Section:Rossby}

\subsection{Definition}
\label{def:rossby}

The Rossby number Ro was first defined by \citet{Rossby39} as the ratio of advection to the Coriolis acceleration. The advective term can be expressed as $\sim U^2/L$ 
(which gives the amplitude of $\left(\vec u\cdot\vec\nabla\right)\vec u$) where $U$ and $L$ are the characteristic speed and length scale of the system, 
respectively, and the Coriolis term is $\sim U\Omega$, with $\Omega$ representing the angular velocity of the star. The Rossby number is thus 
given by 
\begin{equation}
Ro \sim \frac{U^2}{L} \frac{1}{\Omega U} = \frac{U}{L\Omega}.
\end{equation}
In the case of a star, the ratio $U/L$ is the characteristic flow timescale of
the stellar fluid, that is, the convective turnover timescale, and $1/\Omega$ is, to a factor $2 \pi$, the characteristic period of the system, that is, the stellar
rotation rate ($P_{rot} = 2\pi/\Omega$), which leads to   
\begin{equation}
Ro = \frac{P_{rot}}{\tau_{c}}.
\end{equation}
A system that is strongly affected by Coriolis forces will then have a small
Rossby number, while a system in which inertial forces dominate will have a large Rossby number.   
Classically, dynamo action is expected to become more efficient when Ro is lower than 1 (e.g., \citealt{Brunetal2015}, \citealt{Augustsonetal16}, and references therein). Indeed, when the Rossby number is small, convective flows are strongly influenced by the Coriolis acceleration. As a consequence, they become quasi-2D and they generate a stronger kinetic helicity coherent in space and in time. This kinetic helicity is a key actor to sustain an efficient $\alpha$-effect that generates a mean poloidal field thanks to the action of small-scale convective vorticies twisting the mean toroidal magnetic field. In addition, a low convective Rossby number supports a strong differential rotation (e.g., \citealt{Brownetal08,Varelaetal16}), which strengthens the $\Omega$-effect where the differential rotation winds up the mean poloidal field into a mean toroidal magnetic field. Therefore, when the Rossby number decreases both $\alpha$ and $\Omega$ effects and the resulting dynamo action are expected to be stronger. This is also predicted thanks to new theoretical scaling laws giving magnetic energy as a function of the convective Rossby number \citep{2017arXiv170102582A}.

In the literature, the Rossby number is generally computed with the use of the
turnover timescale at half a pressure scale height above the base of the
convective envelope $\tau(H_p/2)$, which we call Ro$(H_p/2)$.  However, for giant stars that have deep and extended convective envelopes, it is not clear where the dynamo operates. 
Therefore, we also compute the
various Rossby numbers for the different $\tau_{c}(r)$ within the convective envelope, and we use 
\begin{equation}
P_{rot} = 2\pi \frac{R_*}{V_{surf}}, 
\end{equation}
with $V_{surf}$ the surface velocity of our rotating models.
For core convection (Appendix), we compute Ro(core) using the global turnover timescale $\tau_{g,core}$ and the mean angular velocity of the core.

\subsection{Rossby number in the convective envelope along the evolution and the theoretical magnetic strips}
\label{Rossby}

\subsubsection{The 2~M$_{\odot}$, Z$_{\odot}$ case}
\label{2M:rossby}

\begin{figure}[!ht]
 \centering
  \includegraphics[width=0.45\textwidth]{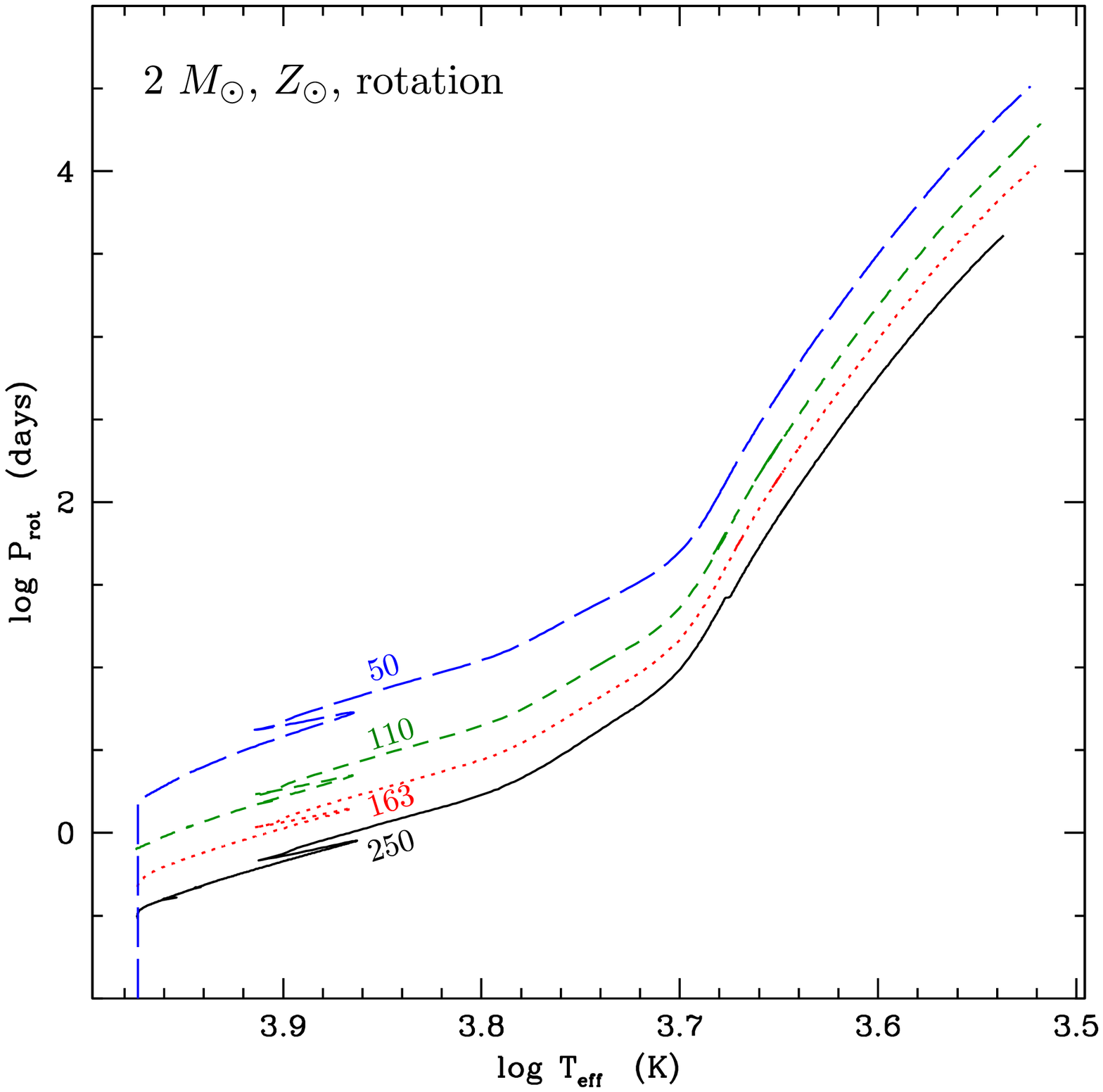}
    \includegraphics[width=0.45\textwidth]{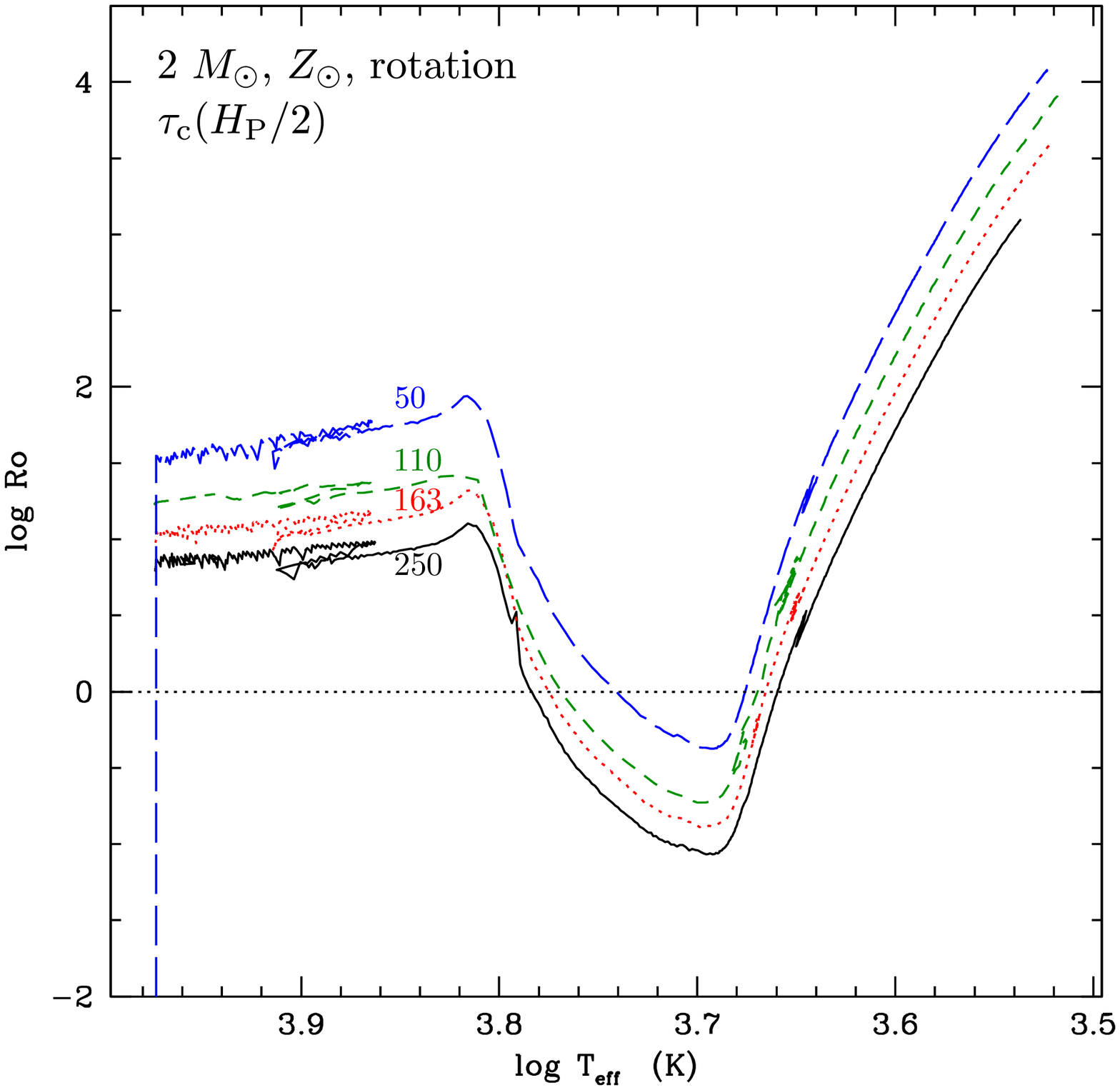}\\  
 \hspace{0.cm} 
  \caption{Evolution of the rotation period (top) and Rossby number computed with the turnover timescale $\tau(H_p/2)$ (bottom)
    from the ZAMS up to the RGB tip for the 2~M$_{\odot}$, Z$_{\odot}$ models computed with different initial rotation velocities on the ZAMS (50, 110, 163, and 250 km sec$^{-1}$). The dotted horizontal line indicates a Ro equal to 1, which is the typical value to invoke $\alpha - \Omega$ dynamo}
  \label{fig:Prot_2MsunZsun}
\end{figure}

Fig.~\ref{fig:Prot_2MsunZsun} shows the evolution of the rotational period for 2~M$_{\odot}$, Z$_{\odot}$ models computed with
different initial rotation velocities, from the ZAMS up to the tip of the RGB.
Since no magnetic braking is applied to this model,
the variations of $P_{rot}$ are driven by secular evolution for this mass range, that is, by the
variations of the stellar radius $R_*$. 

Fig.~\ref{fig:Prot_2MsunZsun} illustrates the ZAMS-to-RGB evolution of
the Rossby number Ro$(H_p/2)$ for the 2~M$_{\odot}$, Z$_{\odot}$ models computed with different initial rotation velocities.  Ro$(H_p/2)$ drops when the convective envelope deepens in mass during the 1DUP. We obtain the same behavior for all Ro calculated with the different convective timescale definitions.
For all the initial rotation rates assumed here, the different Ro 
 than unity, which is the classical value at which 
$\alpha - \Omega$ dynamos are expected to become stronger, although the $T_{eff}$ at which
this occurs during the Hertzsprung gap depends on the initial rotation
velocity (the higher the V$_{ZAMS}$, the earlier Ro reach values 
below unity).
After they reach a minimum at the base of the RGB, all Ro quickly
increase mainly due to the expansion of the envelope as the star climbs the RGB. 
It is interesting to note that for a given mass, Fig.~\ref{fig:Prot_2MsunZsun} predicts that the width of the magnetic strip depends on the initial rotation rate (see  Fig.~\ref{fig:Prot_2MsunZsun} for the evolution of the rotation rates). The faster rotating models have log Ro $<$0 for a wider range of temperatures (respectively $\sim$ 1450 and 700~K along the RGB phase for the faster and the slower rotation rates we assumed for the 2~M$_{\odot}$, Z$_{\odot}$ models). This prediction might be measurable statistically. Finally, at central helium ignition the stellar radius and Ro decrease again. Ro remains of the order of unity while the stars stay on the clump and evolve on the early-AGB before increasing again as the star expands along the TP-AGB phase (not shown in
Fig.~\ref{fig:Prot_2MsunZsun}, but see Fig.~\ref{fig:hrdRossby_compobs_BPI}). 
Therefore, the models predict that the stars cross the magnetic strip twice during their post-main evolution, first on their way to the RGB, and second during central helium burning phase and the early-AGB.

\subsubsection{Mass dependence}
\label{Rossby_all}

\begin{figure}[t]
  \includegraphics[width=0.45\textwidth, angle=0]{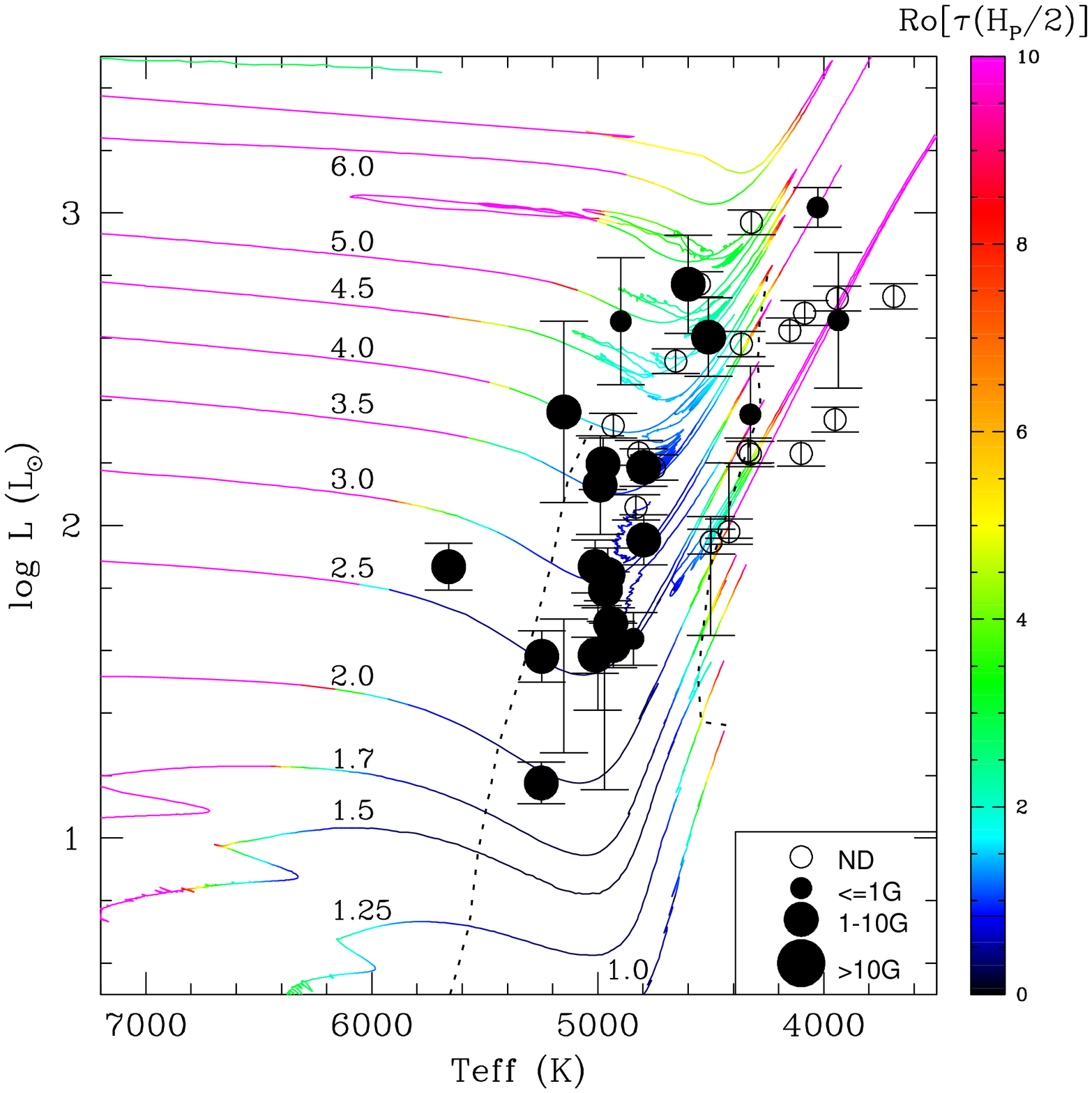}
  \includegraphics[width=0.45\textwidth, angle=0]{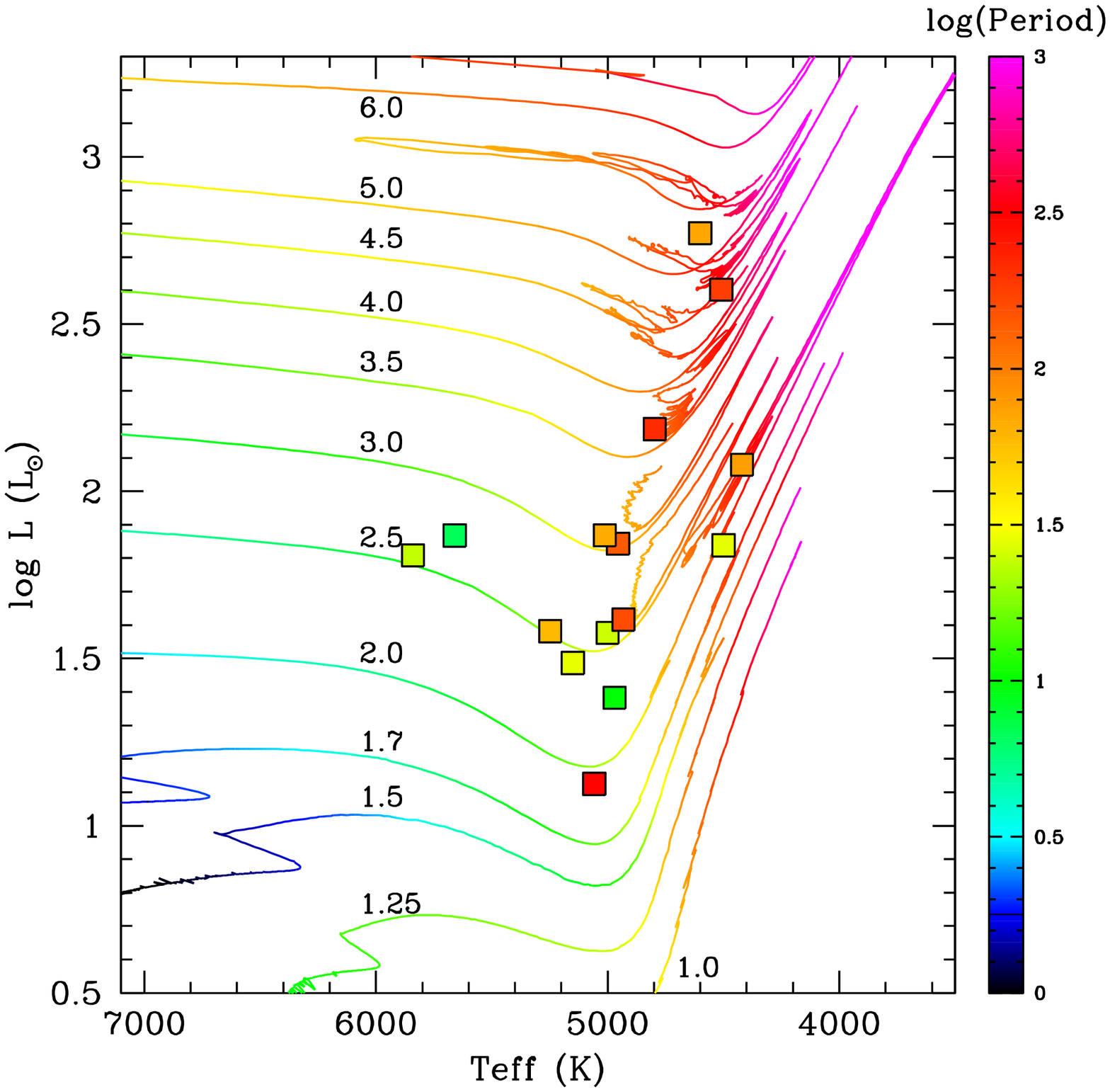}
  \caption{(top) Color-coded Rossby number Ro$(H_p/2)$ along the tracks in the HRD up to the RGB luminosity where Ro$(H_p/2)$ becomes higher than 10 (for stellar masses lower than 2.5~\Ms) or up to the early-AGB (for stellar masses higher than 2.5~\Ms), for the \Zs{} models with rotation. Initial stellar masses are indicated on the tracks. Dotted lines indicate the beginning and end of the first dredge-up (when the mass of the convective envelope is 2.5$\%$ of the total stellar mass, left; at the maximum extent in mass of the convective envelope, right). The points correspond to the sample stars of \citet{Auriereetal2015}; the size of the black symbols depicts the strength of the detected magnetic field, and open symbols stand for nondetected stars. (bottom) A color-coded logarithm scale is shown for the rotation period (in days) along the tracks and for the sample stars of \citet{Auriereetal2015} that have known rotation periods (squares)}
  \label{fig:hrdRossby_compobs_BPI}
\end{figure}

The general behavior of Ro$(H_p/2)$ along the evolution path of our models of
different masses is shown in Fig.~\ref{fig:hrdRossby_compobs_BPI} for the grid at solar metallicity (the comparison with the observations is made in \S~\ref{section:comparison}).
The models with masses lower than or equal to 4~\Ms{}  behave like the 2~\Ms{} case described above. For all of them,
Ro$(H_p/2)$ (as well as Ro$_{max}$ and all Ro(r)) drops below unity
when the stars first cross the Hertzsprung gap and start ascending the RGB.
The Rossby numbers then increase again until central
helium burning ignites.
In this region of the Hertzsprung-Russell diagram, a clear ``magnetic
strip" appears, where Rossby numbers
consistently below unity are predicted and where an $\alpha-\Omega$ dynamo
may thus be stronger.
For higher mass stars the Rossby numbers do not reach such a low value in the
post-MS phases, either because the stellar radius and the rotational period are
higher (\Zs{} case), or because the stars ignite central helium-burning before
they could ascend the RGB. We find the same behavior for all the metallicities.

For the stars with initial masses lower or equal to 4~M$_{\odot}$,
Ro$(H_p/2)$ (as well as all Ro(r)) drops again below unity during central helium-burning, due to the decrease of the
stellar radius.  Rossby numbers rise again when the stars move towards the early-AGB phase.  
This second ``magnetic strip" extends towards higher luminosity and more
advanced evolution phases when the metallicity decreases.
   
In summary, the theoretical magnetic strips we just described appear to be well defined in
terms of effective temperature range (i.e. for a given mass and metallicity this strip is well defined by using only the effective temperature; see Fig. \ref{fig:hrdRossby_compobs_BPI}).  
They comprise stars at the base of the RGB and in the core-helium burning phase.
There appears to be an upper luminosity limit (around Log L/L$_{\odot} \sim 3$), as
more massive stellar models (at Z$_\odot$) or models of lower metallicity
that exhibit blue loops during
central helium burning do not predict Ro(r) decreasing below unity.

\subsubsection{Short note on the case of hot intermediate-mass stars}
Fig.~\ref{fig:hrdRossby_compobs_BPI} reveals our most massive models 
predict $Ro(H_p/2)$ of the order of or lower than unity during some evolutionary
phases at which the convective envelope is extremely thin (e.g., the 6~\Ms,
\Zs{} model on the MS, and the 6~\Ms, Z=0.0001 model from the ZAMS up to the end
of central helium-burning).
These low values of Ro$(H_p/2)$ are mainly due to the high rotation
velocities of the models. However, the subsurface convective layers  in these objects
are caused by the helium opacity bump, and have a very limited extension
(Fig.~\ref{fig:hrdtconvZsunstand}, where all Z$_{\odot}$ models with
initial mass higher than 2~M$_{\odot}$ have convective layers that
are thinner than one pressure scale height); there, convection is very
inefficient and transports only a small fraction of the heat flux.
As a consequence, these stars are not expected to develop $\alpha-\Omega$
dynamo, despite low theoretical Rossby numbers.

\section{Magnetic strips for evolved low- and intermediate-mass stars - Comparison with observations}
\label{section:comparison}

We compare in Fig.~\ref{fig:hrdRossby_compobs_BPI} our model predictions for Ro$(H_p/2)$ along the evolution tracks with the positions in the HRD of the spectro-polarimetric sample of evolved stars studied by \citet{Auriereetal2015}. 
We exclude the stars that are probable Ap descendants, and keep only the stars where the Zeeman effect was detected and which are believed to have $\alpha-\Omega$ dynamos. We also show the sample stars that have no magnetic field detected. While the latest ones were chosen through different selection criteria than the other sample stars, they are clearly not active at the level of the $\alpha-\Omega$ dynamo stars detected in the magnetic strip (that is, below $\sim$ 1~G; see  \citealt{Auriereetal2015} for details). 
Figure~\ref{fig:hrdRossby_compobs_BPI} clearly shows that the vast majority of Zeeman detected evolved stars lie in the area along the tracks where Ro$(H_p/2)$ is close to minimum values around or below unity, while most of the non-magnetic stars lie outside the theoretical magnetic strips (that is, the areas where Ro$(H_p/2)$ becomes much larger than unity).  The only exceptions are the couple of very bright, very cool giant stars with weak detected magnetic fields (below 1~G) for which we predict a Rossby number of the order of 10; unfortunately, the rotation period of these two stars is unknown, but they have very low Vsini, indicating that the $\alpha-\Omega$ dynamo is unlikely to operate in these objects. 

We remind that the values of Ro$_{max}$ (that is, computed with $~\tau_{max}$) along individual tracks are lower than the values of Ro$(H_p/2)$. Since it is not well established yet where the dynamo operates in the convective envelope of cool evolved stars, the agreement seen in Fig.~\ref{fig:hrdRossby_compobs_BPI} between the theoretical and observational magnetic strips appears thus to be very satisfactory. 
This is remarkable especially given that our stellar models are computed for only one initial rotation rate (30$\%$ of the critical rotation velocity on the ZAMS).
However and as can be seen in Fig.~\ref{fig:hrdRossby_compobs_BPI}, the observations reveal a spread in rotation periods for stars that lie close to one another in the Hertzsprung gap, which probably reflects the fact that the sample stars may have started their life with a range of initial rotation velocities. 
This effect can be estimated from Fig.~\ref{fig:Prot_2MsunZsun}, where we show the evolution of the theoretical rotation period for the 2~M$_{\odot}$, Z$_{\odot}$ model computed with different initial rotation rates.  It is dominated by the fast secular increase of the stellar radius that occurs much faster than the internal transport of angular momentum when the star is crossing the Hertzsprung gap.
At the effective temperature of one of the lowest-mass stars observed by \citet[][ HD 203 387; T$_{\rm{eff}}$ = 5012~K, log L/L$_{\odot}$ = 1.87]{Auriereetal2015} the theoretical period of the 2~M$_{\odot}$ model varies between 10 and 60 days (which implies a factor of 6 variation on the theoretical Rossby between the two models) depending on the initial rotation velocity, while the observed period for this star is 68 days. While the overall agreement is relatively good, comparison with the magnetic and rotational properties of individual stars requires tailor-made models with different initial rotation rates for all individual sample stars \citep[e.g.,][]{Auriereetal2009,Auriereetal2011,Auriereetal2012,Konstantinovaetal2010,Konstantinovaetal2012,Tsvetkova_etal13,Borisovaetal16,Tsvetkovaetal16arXiv}, which is out of the scope of the present paper. 

According to our models, the magnetic strips we observe and predict for intermediate-mass stars should extend to the low-mass, low-luminosity regime.  Low-mass stars (below 2~M$_{\odot}$) should thus also undergo $\alpha-\Omega$ dynamo mechanisms while they cross the Hertzsprung gap and reach the base of the RGB\footnote{We can not compute the theoretical Rossby numbers at the clump and on the AGB for the low-mass models of \citet{Lagardeetal12}, since 
they do not follow rotation beyond the RGB tip.}. This could easily be tested observationally, although to the best of our knowledge, there is no magnetic field detection in low-mass (that is, below 1.5~M$_{\odot}$) RGB stars quoted in the literature. 

\section{Conclusions}
\label{conclusion}

Both quantities used to compute the Rossby number, that is, convective and rotational velocities, depend sensitively on the stellar mass, metallicity, and evolutionary phase.
Here we use stellar evolution models that include rotation (Lagarde et al. 2012) to predict these quantities and their temporal variation due to stellar evolution for a large range of initial masses and metallicities. We compute convective turnover times and Rossby numbers at different depths in the stellar convective envelope. 
Although the predicted absolute values for these quantities depend on model input physics, such as $\alpha_{MLT}$ (e.g., \citealt{KimDemarque1996}) and initial rotation rate, the evolutionary (temporal) trends are expected to be largely insensitive to such assumptions. Thus, the predicted tendencies for small Rossby numbers during certain evolutionary phases should be robust, providing a likely explanation of recent magnetic field detections in giant stars in terms of an $\alpha-\Omega$ dynamo mechanism analogous to solar-like stars.

Our models reveal sharp increases in turnover timescales and decreases in rotation rate that are associated to secular evolution effects. This translates into sharp decreases in Rossby number at certain evolutionary phases. These phases show up as well-delineated regions in the HRD that correspond to the first dredge-up phase at the base of the RGB, and to core-helium burning and early AGB. The predicted low Rossby numbers suggest that $\alpha-\Omega$ dynamos may operate in the convective envelopes of low- and intermediate-mass stars during these phases. This offers an explanation for the observed magnetic strips in the HRD that are populated by evolved intermediate-mass stars exhibiting strong magnetic features \citep{Konstantinovaetal2013,Auriereetal2015}. Our prediction that post-main sequence low-mass stars  also cross the magnetic strips could be easily tested in the near future with dedicated spectro-polarimetric observations.  Our models also predict that the width (in effective temperature) of the magnetic strips depends on the initial rotation rate. Thus, the dispersion of Teff among magnetic giants with fast rotation should be statistically larger than that among slower rotators. This could in principle be tested for a statistically significant sample of evolved stars with magnetic detection and rotation rate determination.
Finally, these predictions provide guidance for future magnetohydrodynamic simulations of global stellar dynamos aimed at understanding the precise locus of magnetic field generation in the outer convective layers of evolved stars.

\section*{Appendix - Turnover timescale and Rossby number in stellar convective cores along the main sequence} \label{appendix} 
\label{section:coreconvection}
\begin{figure}
\centering
  \includegraphics[width=0.37\textwidth, angle=-90]{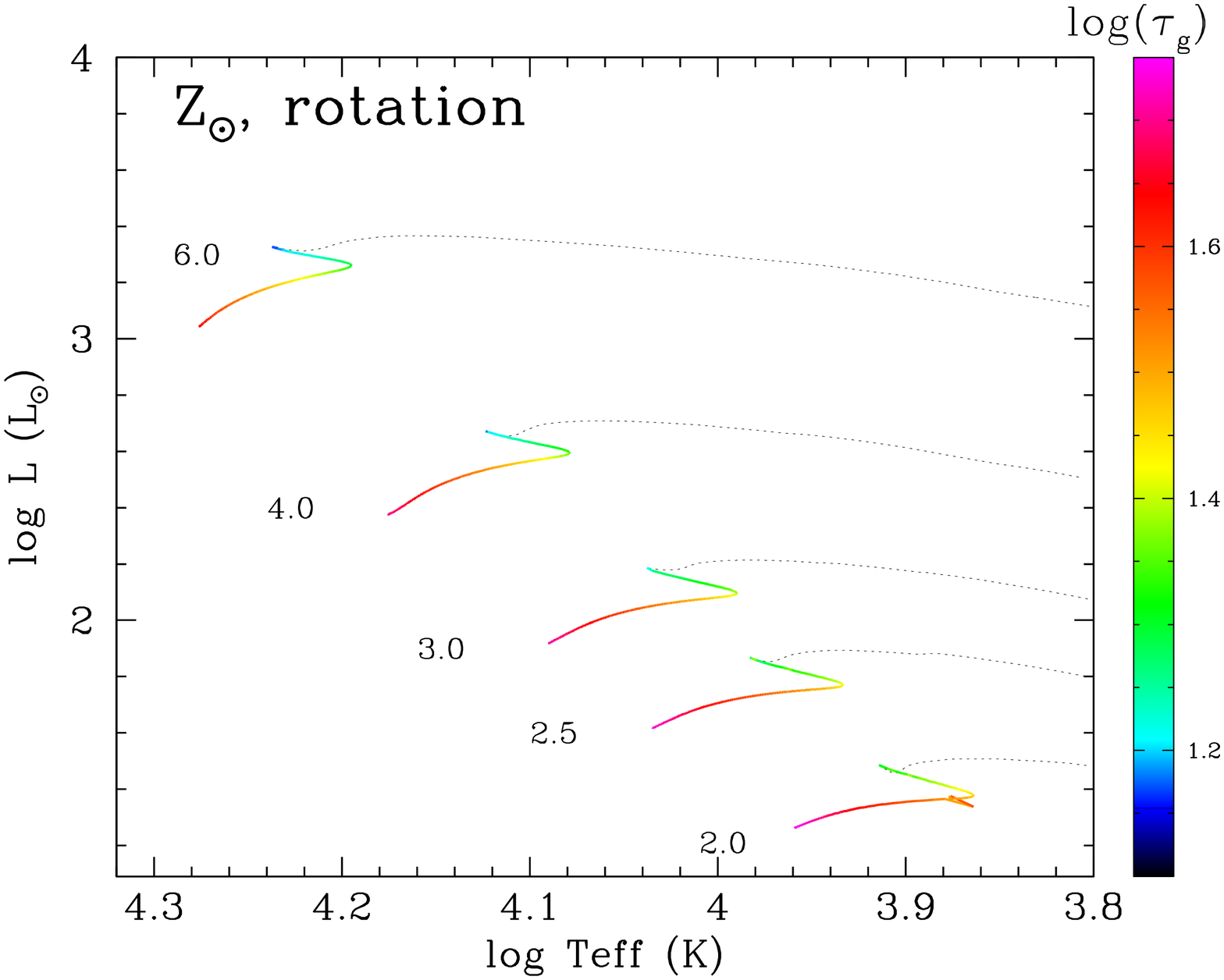} \hspace{0.cm}  
  \includegraphics[width=0.37\textwidth, angle=-90]{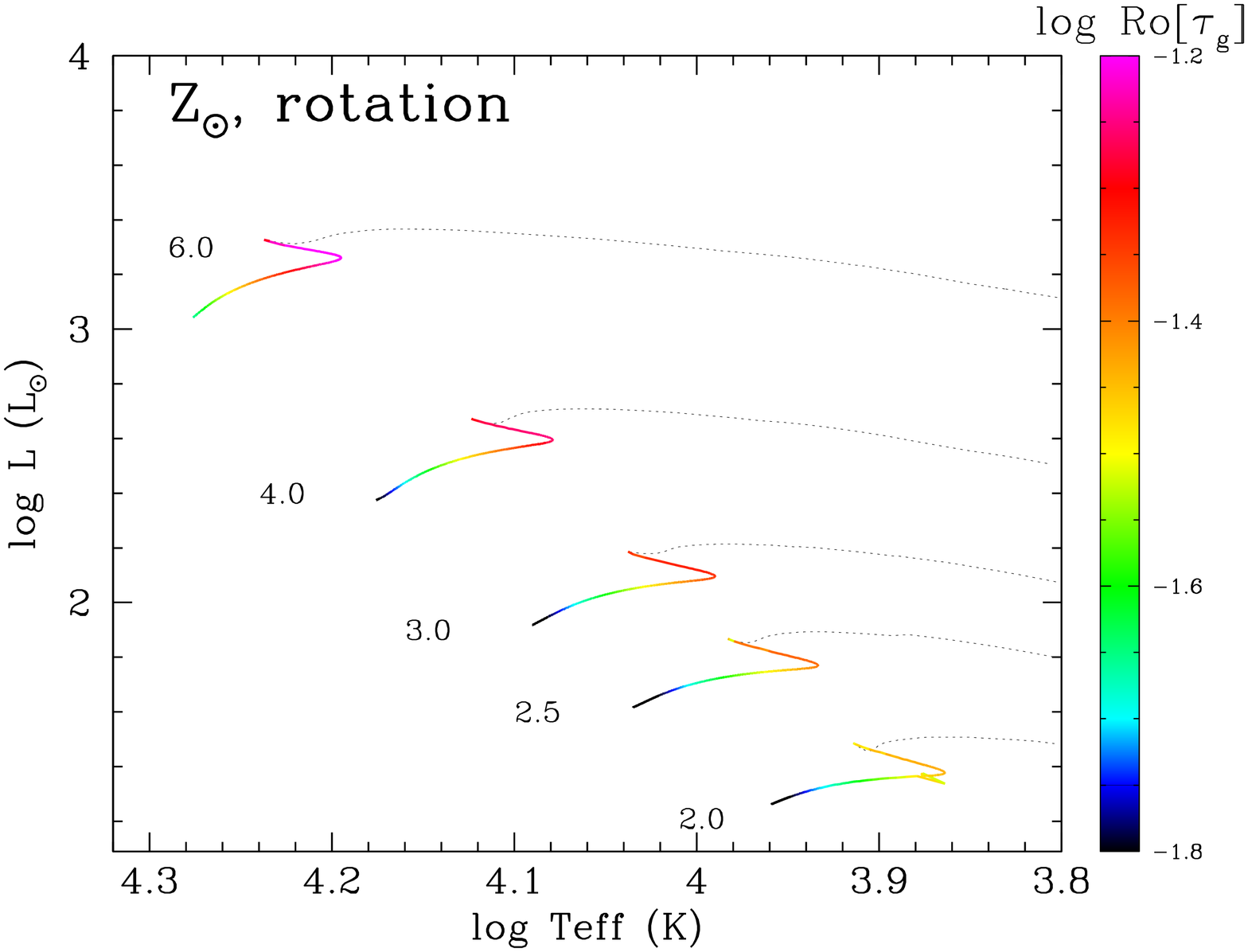} \hspace{0.cm}
\caption{Color-coded global convective turnover time $\tau_g$ and corresponding Rossby number (in logarithmic scale) in the stellar convective core along the MS for the 
solar metallicity models. Initial stellar masses (in solar mass) are indicated on the tracks}
  \label{fig:hrdtconvcoreZsunstand}
\end{figure}

This paper focusses on possible development of $\alpha-\Omega$ dynamo along the so-called magnetic strips in the advanced phases of stellar evolution \citep[][]{Konstantinovaetal2013,Auriereetal2015} when stars have extended convective envelopes. However, internal magnetic fields might also be produced in intermediate-mass main sequence stars that host convective cores \citep[e.g.,][]{Brunetal2005,Stelloetal2016Nature}. Therefore, we have computed the turnover timescales and Rossby numbers in stellar convective cores along the main sequence for Z$_{\odot}$ models with masses between 2 and 6~M$_{\odot}$. We show in Fig.~\ref{fig:hrdtconvcoreZsunstand} the global turnover timescale for the entire core and the associated Rossby number. 
 As for the case of the convective envelope, the values we get for the turnover timescale in the convective core are not affected by rotation, and their quantitative estimate is solid (within the MLT framework). For all the models $\tau_g$  decreases in the core along the MS (for the 2~M$_{\odot}$, Z$_{\odot}$ model, $\tau_g$ is very similar in the convective core and the thin convective envelope; compare Fig.~\ref{fig:m2p0zsunZ0001}). 
On the other hand, the rotation rate of the convective core increases along the MS for all the models because of secular evolution effects. 
The Rossby number of the core thus also slightly increases with time, and this evolution trend is a robust prediction. For the present models, the Rossby number of the core remains well below unity all along the MS. 
This results from the relatively weak coupling between the central and the external regions in the models of \citet{Lagardeetal12}, due to the adopted assumptions for the transport of angular momentum by turbulence and meridional circulation. However, the theoretical core rotation rates are overestimated compared to asteroseismic constraints, which is a well known problem for all the current models including rotation \citep{TZ1998,TCetal2010,Denissenkovetal10,Eggenetal2012,Beck2012,Deheuvels2012,Mosser2012b,Deheuvelsetal2014,Marquesetal2013,Ceillieretal2013,DiMauro2016}. 
Therefore, the values we derive for the central Rossby number are certainly underestimated. 
A consistent study of the dynamo in convective stellar cores would thus require models including very efficient transport of angular momentum in stellar interiors, with processes related, for example, to internal gravity waves \citep{TalCha03,TC98}, to fossil magnetic fields  \citep[][]{CMG1993,GM1998,Eggenbergeretal05,SBZ2011}, or including new prescriptions for  rotation-induced turbulence proposed by Mathis et al. (2017a, submitted to A\&A) that lead to better agreement with asteroseismology (Mathis et al., 2017b, submitted to A\&A; Amard et al., 2017, submitted to A\&A).  This is out of the scope of the present paper.

\begin{acknowledgements}
  The authors acknowledge financial support from the Swiss National Science Foundation (SNSF) and the French Programme National National de
  Physique Stellaire (PNPS) of CNRS/INSU. 
  TD acknowledges financial support from the UE Programme (FP7/2007-2013)
under grant agr. No. 267251 "Astronomy Fellowships in Italy" (ASTROFit). NL  acknowledges financial support from Marie Curie Intra-european fellowship (FP7-PEOPLE-2012-IEF) and from the CNES fellowship. RK-A acknowledges the support of the Bulgarian-French program RILA-CAMPUS, project DRILA 01/3. FG acknowledges financial support from COST Action TD 1308. S.M. acknowledges funding by the European Research
Council through ERC grant SPIRE 647383. We thank the anonymous referee for constructive comments that helped improving the presentation of our results.
\end{acknowledgements}

\bibliographystyle{aa}
\bibliography{Charbonneletal17_Reference}

\end{document}